%% file: main.tex
\documentclass[sigconf]{acmart}
\usepackage{xspace,balance,tabularx,multirow}
\usepackage{flushend}
\usepackage{tikz}
\usepackage{pgfplots}
\usetikzlibrary{patterns}
\pgfplotsset{compat=1.16}
\usepackage{subcaption}
\usepackage[ruled, vlined, linesnumbered]{algorithm2e}
\usepackage{xcolor}
\usepackage{float}
\usepackage{colortbl}
\usepackage{hyperref}
\usepackage{bbold}
\usepackage{pifont}
\SetKwComment{Comment}{$\triangleright$\ }{}
\usepackage{enumitem}
\usepackage[aboveskip=-4pt]{subcaption}

\AtBeginDocument{%
  \providecommand\BibTeX{{%
    \normalfont B\kern-0.5em{\scshape i\kern-0.25em b}\kern-0.8em\TeX}}}

\copyrightyear{2022} 
\acmYear{2022} 
\setcopyright{rightsretained}
\acmConference[CIKM '22] {Proceedings of the 31st ACM International Conference on Information and Knowledge Management}{October 17--21, 2022}{Atlanta, GA, USA.}
\acmBooktitle{Proceedings of the 31st ACM International Conference on Information and Knowledge Management (CIKM '22), October 17--21, 2022, Atlanta, GA, USA}
\acmPrice{}
\acmISBN{978-1-4503-9236-5/22/10} 
\acmDOI{10.1145/3511808.3557076
}

\input{cmd.tex}

\begin{document}

\title{Measuring Friendship Closeness: A Perspective of Social Identity Theory}


\author{Shiqi Zhang}
\authornote{This work was done while Shiqi Zhang was an intern at Tencent.}
\affiliation{%
  \institution{National University of Singapore}
  \city{Singapore}
  \country{Singapore}
}
\affiliation{%
  \institution{Southern University of Science and Technology}
  \city{Shenzhen}
  \country{China}
}
\email{s-zhang@comp.nus.edu.sg}

\author{Jiachen Sun}
\email{jiachensun@tencent.com}
\author{Wenqing Lin}
\authornote{Wenqing Lin and Bo Tang are the corresponding authors.}
\email{edwlin@tencent.com}
\affiliation{%
  \institution{Tencent Inc.}
  \city{Shenzhen}
  \country{China}
}


\author{Xiaokui Xiao}
\affiliation{%
  \institution{National University of Singapore}
  \city{Singapore}
  \country{Singapore}
}
\email{xkxiao@nus.edu.sg}

\author{Bo Tang}
\authornotemark[2]
\affiliation{%
  \institution{Southern University of Science and Technology}
  \city{Shenzhen}
  \country{China}
}
\email{tangb3@sustech.edu.cn}

\begin{abstract}
Measuring the closeness of friendships is an important problem that finds numerous applications in practice. For example, online gaming platforms often host friendship-enhancing events in which a user (called the source) only invites his/her friend (called the target) to play together. In this scenario, the measure of friendship closeness is the backbone for understanding source invitation and target adoption behaviors, and underpins the recommendation of promising targets for the sources. However, most existing measures for friendship closeness only consider the information between the source and target but ignore the information of groups where they are located, which renders inferior results. To address this issue, we present new measures for friendship closeness based on the social identity theory (\sit), which describes the inclination that a target endorses behaviors of users inside the same group. The core of \sit is the process that a target assesses groups of users as them or us. Unfortunately, this process is difficult to be captured due to perceptual factors. To this end, we seamlessly reify the factors of \sit into quantitative measures, which consider local and global information of a target's group. We conduct extensive experiments to evaluate the effectiveness of our proposal against 8 state-of-the-art methods on 3 online gaming datasets. In particular, we demonstrate that our solution can outperform the best competitor on the behavior prediction (resp. online target recommendation) by up to 23.2\% (resp. 34.2\%) in the corresponding evaluation metric.
\end{abstract}



\maketitle

\input{intro.tex}

\input{prelim.tex}

\input{overview.tex}

\input{metric.tex}
\input{exp.tex}

\input{others.tex}

\begin{acks}
This work was supported by Proxima Beta (Grant No.\ A-8000177-00-00) and the Guangdong Provincial Key Laboratory (Grant No.\ 2020B121201001).
Bo Tang is also affiliated with the Research Institute of Trustworthy Autonomous Systems, Shenzhen, China and Guangdong Provincial Key Laboratory of Brain-inspired Intelligent Computation, China.
\end{acks}


\bibliographystyle{ACM-Reference-Format}
\balance
\bibliography{main}


\end{document}

%% file: cmd.tex
\newcommand{\argmin}[1]{\underset{#1}{\operatorname{arg}\,\operatorname{min}}\;}

\makeatletter
\newcommand*\bigcdot{\mathpalette\bigcdot@{.5}}
\newcommand*\bigcdot@[2]{\mathbin{\vcenter{\hbox{\scalebox{#2}{$\m@th#1\bullet$}}}}}
\makeatother

\SetKw{Break}{break}

\newcommand{\ie}{{\it i.e.},\xspace}
\newcommand{\eg}{{\it e.g.},\xspace}

\newcommand{\stitle}[1]{\vspace*{0.5em}\noindent{\bf #1.\/}}

\newtheorem{problem}{Problem}

\newcommand{\codmrecall}{{\em FPS}\xspace}
\newcommand{\lgamerecall}{{\em MOBA-A}\xspace}

\newcommand{\xinshi}{{\em MOBA-B}\xspace}

\newcommand{\Gset}{\mathcal{G}\xspace}
\newcommand{\V}{\mathcal{V}\xspace}
\newcommand{\E}{\mathcal{E}\xspace}

\newcommand{\R}{\mathcal{R}\xspace}
\newcommand{\N}{\mathcal{N}\xspace}
\newcommand{\Sset}{\mathcal{S}\xspace}
\newcommand{\Tset}{\mathcal{T}\xspace}
\newcommand{\Cset}{\mathcal{C}\xspace}

\newcommand{\tfc}{${\mathtt{TFC}}$\xspace}

\newcommand{\cc}{${\mathsf{CC}}$\xspace}
\newcommand{\ccs}{${\mathsf{CCs}}$\xspace}
\newcommand{\igt}{${\mathsf{IGT}}$\xspace}
\newcommand{\ugt}{${\mathsf{UGT}}$\xspace}
\newcommand{\gs}{${\mathsf{GS}}$\xspace}
\newcommand{\cpr}{${\mathsf{GPR}}$\xspace}
\newcommand{\cppr}{${\mathsf{GPPR}}$\xspace}

\newcommand{\ppr}{${\mathsf{PPR}}$\xspace}

\newcommand{\sit}{${\mathsf{SIT}}$\xspace}
\newcommand{\sitb}{$\boldsymbol{\mathsf{SIT}}$\xspace}
\newcommand{\ntovcos}{${\mathsf{N2V}}$(${\mathsf{cos}}$)\xspace}
\newcommand{\ntoveuc}{${\mathsf{N2V}}$(${\mathsf{euc}}$)\xspace}
\newcommand{\com}{${\mathsf{COM}}$\xspace}
\newcommand{\gt}{${\mathsf{GT}}$\xspace}
\newcommand{\gd}{${\mathsf{GD}}$\xspace}
\newcommand{\tie}{${\mathsf{Tie}}$\xspace}
\newcommand{\xgboost}{${\mathsf{XGBoost}}$\xspace}

\newlength\lengtha 

\definecolor{forestgreen}{RGB}{34, 139, 34}

\definecolor{RYB1}{RGB}{192, 128, 255}
\definecolor{RYB2}{RGB}{255, 192, 32}
\definecolor{RYB3}{RGB}{139, 0, 0}
\definecolor{RYB4}{RGB}{0, 128, 255}

\newenvironment{customlegend}[1][]{%
    \begingroup
    \csname pgfplots@init@cleared@structures\endcsname
    \pgfplotsset{#1}%
}{%
    \csname pgfplots@createlegend\endcsname
    \endgroup
}%

\def\addlegendimage{\csname pgfplots@addlegendimage\endcsname}

\makeatletter
\newcommand\footnoteref[1]{\protected@xdef\@thefnmark{\ref{#1}}\@footnotemark}
\makeatother

\let\oldnl\nl
\newcommand{\nonl}{\renewcommand{\nl}{\let\nl\oldnl}}

%% file: intro.tex
\section{Introduction}\label{sec:intro}
Consider a social network $\Gset=(\V,\E)$, where each node $v_s \in \V$ represents the user $v_s$ and each edge $(v_s, v_t) \in \E$ represents the friendship of two users $v_s$ and $v_t$.
Measuring the closeness of the friendship $(v_s,v_t)$ is an important problem and finds numerous applications in real-world scenarios, where a user $v_s$ (called the source) \textit{only} interacts with the friend $v_t$ (called the target).
For example, online gaming platforms often host friendship-enhancing events, which encourage the source $v_s$ to invite the target $v_t$ to return to the game or play together~\cite{luo2019efficient,luo2020efficient}. In this scenario, the friendship closeness measure is the backbone for understanding the invitation (resp. adoption) behavior of $v_s$ (resp. $v_t$). Furthermore, this measure can be applied in recommending promising targets for the sources because the number of targets of a source could be considerable. 
Analogously, in the instant messaging platform WeChat, a source's post can only be liked and commented on by the target in contacts~\cite{zhang2018mobile}, thus the friendship closeness measure can also be leveraged to predict corresponding activities~\cite{zhang2021understanding,sankar2020inf,gomez2011uncovering}.

In the present work, we focus on the \textit{topological friendship closeness} (\tfc) measure that reflects the closeness of a source-target pair $(v_s,v_t)$ on $\Gset$ rather than labels associated with them. In particular, most existing \tfc measures~\cite{adamic2003friends,page1999pagerank,jeh2002simrank,granovetter1978threshold,fang2014modeling,perozzi2014deepwalk,grover2016node2vec,yang13homogeneous,roweis2000nonlinear,tang2015line,wang2016structural} adopt the structural information between two users $v_s$ and $v_t$. For example, \cite{fang2014modeling} directly employs the tie strength of $(v_s,v_t)$, which is measured by the times of historical interactions; \cite{adamic2003friends} proposes several measures in terms of the common neighborhood of $(v_s,v_t)$. Furthermore, some measures preserve long and intricate paths between $(v_s,v_t)$ based on random walks, \eg personalized PageRank~\cite{page1999pagerank}. 
Unfortunately, the above-said measures may render compromised results, as they ignore the group information and the potential \textit{group effect} related to $(v_s,v_t)$, which says a user's belonging group can influence his/her decision. 



Even though some \tfc measures~\cite{soundarajan2012using,epasto2015ego,ugander2012structural,song2020you,barbieri2014follow,de2016discriminative} attempt to take the group effect into consideration, the involved information is still inadequate.
To explain, we find that these measures typically require (i) first representing the affiliation relationship between users and groups as a bipartite graph, where users and groups are two disjoint sets of nodes and affiliation relationships are a set of edges, and (ii) then exploiting information on this bipartite graph. For example, \cite{ugander2012structural} proposes structural diversity to measure the number of adjacent groups for a given user; \cite{song2020you,barbieri2014follow,de2016discriminative} utilize edge weights between users and groups; \cite{soundarajan2012using,epasto2015ego} measures the commonality of groups between two users on this bipartite graph. However, in these works, contracting a group into a node makes more detailed group information imperceptible, \eg connectivity.

To mitigate the deficiencies of existing measures, we propose to explore the group effect by leveraging the social identity theory (\sit), which is a fundamental concept in social psychology and is widely applied in domains of human health~\cite{scheepers2019social}, team sports~\cite{rees2015social}, computer-supported cooperative works~\cite{seering2018applications} and fake news detection~\cite{shu2017fake}. 
\sit describes that a target $v_t$ tends to endorse attitudes and behaviors of groups of users, who are assessed as us in $v_t$'s cognition.
In other words, given a target $v_t$ and an identified group $\Cset$ of $v_t$, $v_t$ is more likely to adopt the invitation from the source $v_s\in \Cset$. Unfortunately, the identification process is difficult to be quantified due to perceptual factors.
To this end, we seamlessly reify the factors of \sit into \tfc measures, which consider a group's local and global information. For example, by modeling the social network as a physical system, we propose to measure the attractive spring-like force related to a group, in which the information of group connectivity and in-group user homogeneity are naturally incorporated. 
In contrast, previous \tfc measures only consider similarities of group memberships~\cite{soundarajan2012using,epasto2015ego}. 

We experimentally evaluate the proposed \sit-based solution against 8 representative competitors on 3 real-world online game datasets.
In particular, we demonstrate that the proposed solution outperforms all competitors in terms of AUC, accuracy, and F1 score while predicting target adoption and source invitation behaviors on the tested dataset. Besides, we detailedly analyze the relations between user behaviors and \sit factors, and the importance of these factors. At last, we deploy our solution to the online target recommendation, which achieves up to 34.2\% improvement over the best treatment in the corresponding evaluation metric.  

To summarize, we make the following contributions in this work:
\begin{itemize}[topsep=2pt,itemsep=1pt,parsep=0pt,partopsep=0pt,leftmargin=11pt]
    \item We propose new \tfc measures based on \sit in social psychology, which preserve both local and global information of groups.   
    \item We conduct extensive experiments and analysis to demonstrate the superiority of proposed \sit-based \tfc measures over state-of-the-art competitors while predicting user behaviors. 
    \item We deploy the presented solution to the online target recommendation, which achieves significant improvement.
\end{itemize}

%% file: prelim.tex
\section{Preliminaries}\label{sec:prelim}
In this section, we first elaborate on the problem of measuring the topological friendship closeness (\tfc), followed by the illustration of main competitors and downstream tasks that \tfc measures are to solve in this work. At last, we introduce the background of the social identity theory.

\subsection{Problem Formulation}\label{sec:formulation}

Let $\Gset=(\V,\E)$ be a social network, where $\V$ is a set of users (called nodes) and $\E$ is a set of friendships (called edges).
We assume that the friendship $(v_s,v_t)\in \E$ is directed and is associated with an edge weight $w_{s,t}\in (0,1]$. 
Given a directed edge $(v_s,v_t)\in \E$,  we call $v_s$ (resp. $v_t$) the \textit{source} (resp. \textit{target}) neighbor of $v_t$ (resp. $v_s$). 
Given a node $v_t \in \V$, the \textit{source neighborhood} of $v_t$ consists of: (i) $v_t$ itself; (ii) the node set $\N_t$, which contains sources of $v_t$; (iii) the edge set $\E_t$, which contains the edge $(v_s,v_t)\in \E$; (iv) the edge set $\R_t$, which contains the edge $(v_s,v_k)\in \E$ between sources $v_s,v_k\in \N_t$.

Recall in Section~\ref{sec:intro} that, given an input social network $\Gset$, the \tfc measure reflects the topological closeness for each $(v_s,v_t)\in \E$.
Given a source-target pair $(v_s,v_t)\in \E$, the goal of our present work is to design the \tfc measure, which leverages the structural information of the group $\Cset$, where $v_s,v_t\in \Cset$. For simplicity, we assume that $v_s$ and $v_t$ are only co-located in one group $\Cset$.
As for the partitioning of groups, it relies on the social identity theory and will be illustrated in Section~\ref{sec:overview}.

\subsection{Main Competitors}\label{sec:exist-measure}
In what follows, we briefly elaborate on the main ideas of competing \tfc measures in the present work. For better clarity, we call a measure as the \textit{group-level} \tfc measure if considering the group information, and the \textit{individual-level} \tfc measure otherwise.
Specifically, we select 5 representative individual-level measures: tie strength~\cite{tang2015line,fang2014modeling}, number of common neighbors~\cite{tang2015line,adamic2003friends}, personalized PageRank~\cite{page1999pagerank}, cosine and euclidean similarity of Node2Vec vectors~\cite{grover2016node2vec}; 3 commonly-used group-level measures: structural diversity~\cite{ugander2012structural}, user-group tie strength~\cite{song2020you} and group edge density~\cite{purohit2014understanding}. 
It is worth noting that the selected individual-level measures are also known as proximities~\cite{cai2018comprehensive,tang2015line}, which can reflect the topological closeness for all node pairs. Furthermore, these \tfc measures are widely applied in Tencent gaming platforms~\cite{luo2019efficient,luo2020efficient}.

\stitle{Tie strength} Given a friendship $(v_s,v_t)\in \E$, the tie strength equals to the edge weight $w_{s,t}$~\cite{tang2015line,fang2014modeling}.

\stitle{Number of common neighbors} Given a friendship $(v_s,v_t)\in \E$, this measure counts the number of common neighbors~\cite{tang2015line,adamic2003friends} of source $v_s$ and target $v_t$, which can reflect the similarity of two node's local structure. 

\stitle{Personalized PageRank} Given a friendship $(v_s,v_t)\in \E$, this measure aims to preserve intricate topological relations between nodes by performing random walks. In particular, the personalized PageRank~\cite{page1999pagerank} from $v_s$ to $v_t$ is defined as the probability that a \textit{random walk with restart} (RWR)~\cite{tong2006fast} originating from $v_s$ stops at $v_t$. The corresponding RWR starts from the source $v_s$ and, in each step, chooses to (i) either terminate at the current node with probability $\alpha$, (ii) or navigate to a random target neighbor of the current node with the remaining $1-\alpha$ probability.

\stitle{Similarity of Node2Vec vectors} As a node embedding method~\cite{cai2018comprehensive}, Node2Vec~\cite{grover2016node2vec} proposes to represent each node $v_s\in \Gset$ as a compact vector, which preserves the structural information in the vicinity of $v_s$. By this method, given a friendship $(v_s,v_t)\in \E$, the \tfc can be measured by the cosine similarity or euclidean distance between the Node2Vec vectors of $v_s$ and $v_t$. Compared with the prior work DeepWalk~\cite{perozzi2014deepwalk}, the core contribution of Node2Vec is that it takes \textit{two-order random walks} as inputs for training. Compared with the RWR, the two-order random walk will not terminate until a pre-defined length limit. Moreover, given the current node $v_j$, the target neighbors of $v_j$ are first divided into the following three categories: (i) the source neighbor $v_i$ that was visited in the previous step, (ii) $v_k$ that is the common target neighbor of $v_i$ or (iii) otherwise. Then different (resp. same) transition probabilities are assigned to nodes across (resp. inside) each category. At last, the next node of this walk is selected in terms of the biased probability.

\stitle{Structural diversity}
Ugander et al. are the first to propose the concept of structural diversity~\cite{ugander2012structural}, which is further applied for influence analysis~\cite{fang2014modeling,zhang2021understanding,su2020experimental,qiu2016lifecycle}. Specifically, given a friendship $(v_s,v_t)\in \E$, the \tfc measure based on the structural diversity equals to the number of weakly connected components (see Definition~\ref{def:cc}) in a subgraph, which is derived from $v_t$'s source neighbors affected by a specific event and their relationships. This work \cite{ugander2012structural} points out that a more extensive structural diversity of $v_t$ indicates a higher probability that $v_t$ will be influenced. 
\begin{definition}[Weakly Connected Component]\label{def:cc}
Given a directed graph, a weakly connected component is a maximal subgraph, where any inside node is connected by paths if ignoring the direction of edges.
\end{definition}

\stitle{User-group tie strength}
Given a friendship $(v_s,v_t)\in \E$ and a group $\Cset$ where $v_s,v_t\in \Cset$, the \tfc measure based on the user-group tie strength~\cite{song2020you,barbieri2014follow,de2016discriminative} equals to the summation of tie strength between $v_t$ and other members in $\Cset$. 

\stitle{Group edge density}
Given a friendship $(v_s,v_t)\in \E$ and a group $\Cset$ where $v_s,v_t\in \Cset$, the \tfc measure based on the group edge density is defined as the fraction of the summation of edge weights in $\Cset$ over the number of all possible edges in $\Cset$, which is also applied to analyze group dynamics~\cite{purohit2014understanding,qiu2016lifecycle}.

\subsection{Downstream Tasks}\label{sec:task}
Before introducing downstream tasks, we first illustrate friendship-enhancing events in Tencent‘s online gaming platforms, whose procedure is as follows.
First, given a social network $\Gset=(\V,\E)$, a source set $\Sset \subseteq \V$ and a target set $\Tset\subseteq \V$ are selected regarding the event demand, \eg recalling inactive targets and stimulating interactions between friends. Then, the event is sent to each source $v_s\in \Sset$, and allows $v_s$ to invite the target neighbor $v_t \in \Tset$ from a feed window with at most $k$ target friends. Once a target $v_t$ adopts the invitation from the source $v_s$, both $v_s$ and $v_t$ will receive virtual gifts as incentives. To summarize, there exist two types of user activities: \textit{source invitation} and \textit{target adoption}. 
In friendship-enhancing events, \tfc measures act as structural heuristics to facilitate the understanding of invitation and adoption behaviors and recommending targets for the sources, which are defined as follows.

\begin{problem}[Behavior prediction]\label{prob:prediction}
Given a graph $\Gset=(\V,\E)$ and a source-target pair $(v_s,v_t)\in \E$, the objective of behavior prediction is to infer the likelihood that source invitation and target adoption behaviors happen between $(v_s,v_t)$.
\end{problem}

\begin{problem}[Target recommendation]\label{prob:recommend}
Given a graph $\Gset=(\V,\E)$, a budget $k$, a source set $\Sset$ and a target set $\Tset$, the objective of target recommendation is to select at most $k$ target neighbors from $\Tset$ for each source user $v_s\in \Sset$, such that the likelihood that both invitation and adoption behaviors happen among all returned source-target pairs are maximized.
\end{problem}

\subsection{Social Identity Theory}\label{sec:social-identity}
\stitle{Formulation}
Turner and Tajfel define social identity as `the individual’s knowledge that he/she belongs to certain social groups together with some emotional and value significance to him of this group membership'~\cite{tajfel2004social}. In other words, social identity theory (\sit) emphasizes how the target user's belonging group affects his/her behavior.
Given a social network $G$ and a target $v_t$, \sit is formulated by three cognitive processes:  \textit{categorization}, \textit{identification} and \textit{comparison}~\cite{tajfel2004social}. Specifically, the categorization is a preprocessing step, which partitions the users of $\Gset$ into several candidate social groups w.r.t $v_t$. In the identification stage, $v_t$ will identify him/herself as a member of a certain candidate group. Finally, the comparison stage says that, for the sake of self-esteem, $v_t$ tends to shift the sense of belonging based on group social standing.

\begin{table}[!t]
\caption{Factors in \sitb.}\label{tab:sit-factor}\vspace{-2mm}
\renewcommand{\arraystretch}{1.1}
\centering
\begin{small}
\resizebox{1\columnwidth}{!}{%
\begin{tabular}{|c|l|}
\hline
\textbf{Factor}                             & \multicolumn{1}{c|}{\textbf{Meaning}}                                  \\ \hline
Multi-membership                   & Number of groups                          \\ \hline
Inclusiveness                      & Number of in-group members                \\ \hline
Satisfaction      & $v_t$'s feelings about being a group member                \\ \hline
Solidarity        & $v_t$'s psychological bond with in-group members               \\ \hline
Centrality        & Importance of a group in $v_t$'s cognition                   \\ \hline
Self-stereotyping & Similarity of $v_t$ and group average in $v_t$'s cognition       \\ \hline
In-group homogeneity               & Similarity within a group \\ \hline
Social standing                    & Social standing of a group \\ \hline
\end{tabular}
}
\end{small}
\vspace{-2mm}
\end{table}

\stitle{Factors}
After pre-partitioning candidate groups for a target $v_t$, both identification and comparison stages depend on many factors. We summarize the factors used in this paper in Table~\ref{tab:sit-factor}.
Specifically, to understand the identification stage, the statistical factors
multi-membership~\cite{sieber1974toward} and inclusiveness~\cite{brewer1996we} are proposed. Besides, Leach et al.~\cite{leach2008group} extensively summarize existing social perceptual factors as: satisfaction~\cite{tajfel2004social,cameron2004three}, solidarity~\cite{cameron2004three,doosje1998guilty}, centrality~\cite{cameron2004three,doosje1998guilty,brewer1996we}, self-stereotyping~\cite{oakes1994stereotyping,simon1992perception} and in-group homogeneity~\cite{oakes1994stereotyping,simon1992perception}.
Regarding the comparison stage, by definition, the social standing of a group is also recognized as a pivotal factor~\cite{tajfel2004social}.


%% file: overview.tex
\begin{figure*}[t]
\centering
\begin{small}
\hspace{8mm}
\subfloat[Input graph]{
\centering
	\includegraphics[width=0.4\columnwidth]{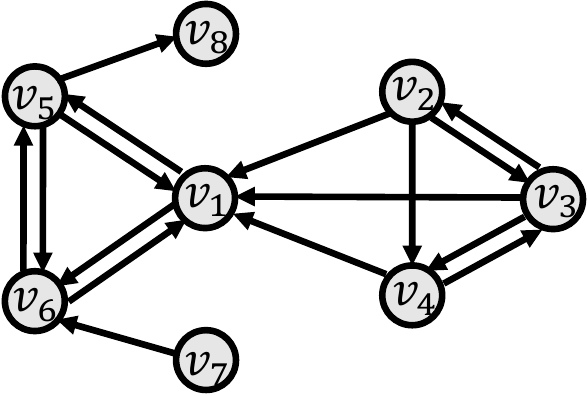}
\vspace{4mm}%
}
\hspace{8mm}
\subfloat[Categorization]{
    \centering
	\includegraphics[width=0.45\columnwidth]{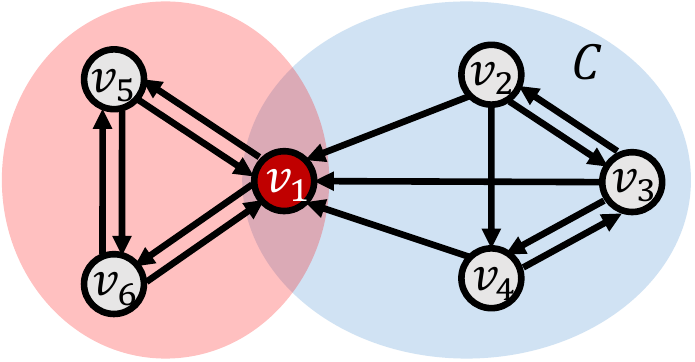}
\vspace{6mm}%
}
\hspace{8mm}
\subfloat[Factor]{
\centering
	\includegraphics[width=0.35\columnwidth]{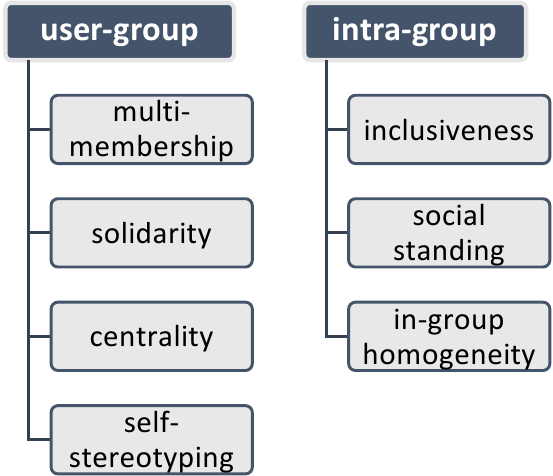}
\vspace{2mm}%
}
\hspace{8mm}
\subfloat[Inference]{
\centering
	\includegraphics[width=0.3\columnwidth]{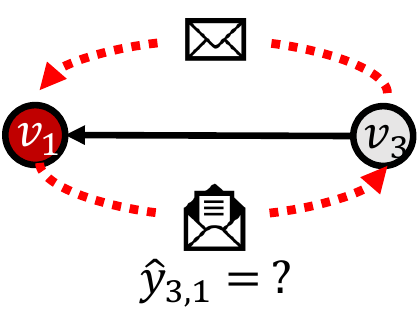}
\vspace{8mm}%
}
\hspace{8mm}
\vspace{-2mm}
\end{small}
\caption{Illustration of core subroutines.}
\vspace{-2mm}
\label{fig:example}
\end{figure*}

\section{Overview}
\label{sec:overview}
We illustrate the main workflow of the present work in Figure~\ref{fig:example}, which processes the input graph by following three steps.

\stitle{Step 1: categorization}
For employing \sit, the first step is partitioning candidate groups for each target user, \ie the categorization stage.
To explore the group information for the source-target pair,
given a graph $\Gset=(\V,\E)$ and a target user $v_t$, we partition the source neighborhood of $v_t$ based on the weakly connected components (see Definition~\ref{def:cc}).
Specifically, following operations in~\cite{ugander2012structural,fang2014modeling,zhang2021understanding,epasto2015ego}, we first extract each weakly connected component (dubbed as \cc) from the ego network $\Gset_t = (\N_t,\R_t)$ of $v_t$, and then call all sources in a given \cc and the target $v_t$ as a candidate group $\Cset$, since $v_t$ connects to all sources in the \cc by edges in $\E_{t}$. To exemplify, given an input graph as shown in Figure~\ref{fig:example}(a), the source neighborhood of target node $v_1$ is partitioned into two candidate groups (shaded in two colors in Figure~\ref{fig:example}(b)), among which the group $\Cset$ is constituted by $v_1,v_2,v_3,v_4$. 
The rationale for employing \cc is two-fold. First, as suggested in \sit~\cite{gaertner2014reducing}, the candidate group requires to be categorized in a more general way to mitigate the in-group bias. For instance, employers in two departments of the same company should be categorized into one company group rather than split into two department groups.
By definition, the \cc can satisfy this requirement, as the inside nodes of two \ccs are isolated (\ie no path).
Second, \cc is a simple but effective way to represent the community in the ego network of a user~\cite{epasto2015ego,yang2015defining}. As evidence, \cite{epasto2015ego} finds that employing \ccs for the link prediction can achieve comparable or even slightly better performance than employing communities detected by the state-of-the-art methods.

\stitle{Step 2: \sitb-based measure definition}
As shown in Figure~\ref{fig:example}(c), factors of \sit in Table~\ref{tab:sit-factor} can be separated into two branches: (i) user-group factors, describing the information between the target and other members; (ii) intra-group factors, describing the stand-alone information of other members. For example, the factors of self-stereotyping and in-group homogeneity both focus on individual similarities, however, by definition, the former is a user-group factor and the latter is an intra-group one.
Accordingly, once the categorization stage is accomplished, given a target $v_t$ and a candidate group $\Cset$, the second step is to devise group-level measures between $v_t$ and $\Cset\backslash v_t$ (resp. among sources in $\Cset\backslash v_t$) to quantify user-group (resp. intra-group) factors of \sit. 
Notice that we skip the satisfaction factor as it involves numerous human sentiments and is difficult to be detected by employing graph topology only.

\stitle{Step 3: Inclination inference}
In the third step, we employ the proposed \sit-based measures to infer the inclination that the target $v_t$ endorses each candidate group $\Cset$. This inclination is defined as $$\hat{y}_{\Cset,t} = \frac{\sum_{v_s\in\Cset\backslash v_t}\hat{y}_{s,t}}{|\Cset|-1},$$ where $|\Cset|\geq 2$ as there exist at least $v_t$ and a source, and $\hat{y}_{s,t}$ is the likelihood that $v_t$ adopts the invitation from each source $v_s$ in $\Cset$ (see Figure~\ref{fig:example}(d)) and can be applied in the downstream Problem~\ref{prob:prediction} and Problem~\ref{prob:recommend}.
Without loss of generality, we derive $\hat{y}_{s,t}$ by a supervised manner, which works as follows. For each training source-target pair $(v_s,v_t)$, we first take the proposed \sit-based measures as $d$-dimensional features $x_{s,t}\in \mathbb{R}^d$ and $y_{s,t}\in \{0,1\}$ as the label of the adoption behavior, where $y_{s,t}=1$ indicates $v_t$ adopts the invitation from $v_s$ and $y_{s,t}=0$ otherwise. We next train a well-accepted \xgboost~\cite{chen2016xgboost} model, which is finally utilized to infer each $\hat{y}_{s,t}$. 
Specifically, given a training dataset with related tuple $(x_{s,t}, y_{s,t})$ and $T$ regression trees with the maximum depth $h$, \xgboost aims to find the best parameters attached on all possible $2^h$ leaves of each of $T$ trees by the following objective where all parameters are represented by a matrix $\mathbf{\Theta}\in \mathbb{R}^{T\times 2^h}$,
\begin{equation}\label{eq:obj}
    \argmin{\mathbf{\Theta}}\left(\mathcal{L}(\mathbf{\Theta})+\Omega(\mathbf{\Theta})\right).
\end{equation}
In Eq.\eqref{eq:obj}, the training loss term is defined as the logistic loss 
\begin{equation}\label{eq:loss}
    \mathcal{L}(\mathbf{\Theta}) = \sum\limits_{s,t}\left(y_{s,t}\ln(1+e^{-\hat{y}_{s,t}})+(1-y_{s,t})\ln(1+e^{\hat{y}_{s,t}})\right),
\end{equation}
and the regularization term $\Omega(\mathbf{\Theta})$ considers both L0 and L2 norms of $\mathbf{\Theta}$. Regarding the predicted value $\hat{y}_{s,t}$ in Eq.\eqref{eq:loss}, it aggregates the parameters by $\hat{y}_{s,t}=\sum_{i}\mathbf{\Theta}[i,q_{i}(s,t)]$, where the $i$-th row of $\mathbf{\Theta}$ contains all parameters in the $i$-th tree and $q_{i}(s,t)$ is the column index that contains the parameter of current $(v_s,v_t)$ in the $i$-th tree.
The reason of employing \xgboost is that (i) it is an empirically-efficient solver~\cite{chen2016xgboost} and is widely accepted for massive game data in Tencent; (ii) the model is easy to be interpreted and is further applied for related analysis in Section~\ref{sec:exp-predict}.

\stitle{Remarks}
In what follows, we focus on defining \sit-based measures in Section~\ref{sec:metric}, where our key contributions come from. The performance of two downstream tasks are evaluated in Section~\ref{sec:exp-predict} and Section~\ref{sec:exp-recommend}.
As selections of community detection method in step 1 and training model in step 3 are orthogonal to our work, we refer the interested readers to ~\cite{epasto2015ego} for the benchmark of different community detection methods and~\cite{chen2016xgboost} for the detailed training procedure of \xgboost, respectively.

%% file: metric.tex
\section{\sitb-based Measures}\label{sec:metric}
In this part, we devise the formulation for \sit factors and illustrate their rationales compared with existing group-level measures.

\subsection{Multi-membership and Inclusiveness}

Inspired from the structural diversity~\cite{ugander2012structural}, we utilize the number of candidate groups and the size of each group of a target to represent the statistical factors of multi-membership and inclusiveness, respectively. In particular, given a graph $\Gset$ and a target user $v_t$, we define the multi-membership of $v_t$ as the number of candidate groups ($\#\mathsf{CC}$) consisting of \textit{all} source neighbors of $v_t$, and the inclusiveness of a given group $\Cset$ as the group size (\gs), \ie the cardinality $|\Cset|$.
In contrast, the prior work of structural diversity~\cite{ugander2012structural} focuses on the groups with source neighbors who have sent the invitation only.

\begin{figure}[t]
\hspace{-4mm}
    \centering
	\includegraphics[width=0.9\columnwidth]{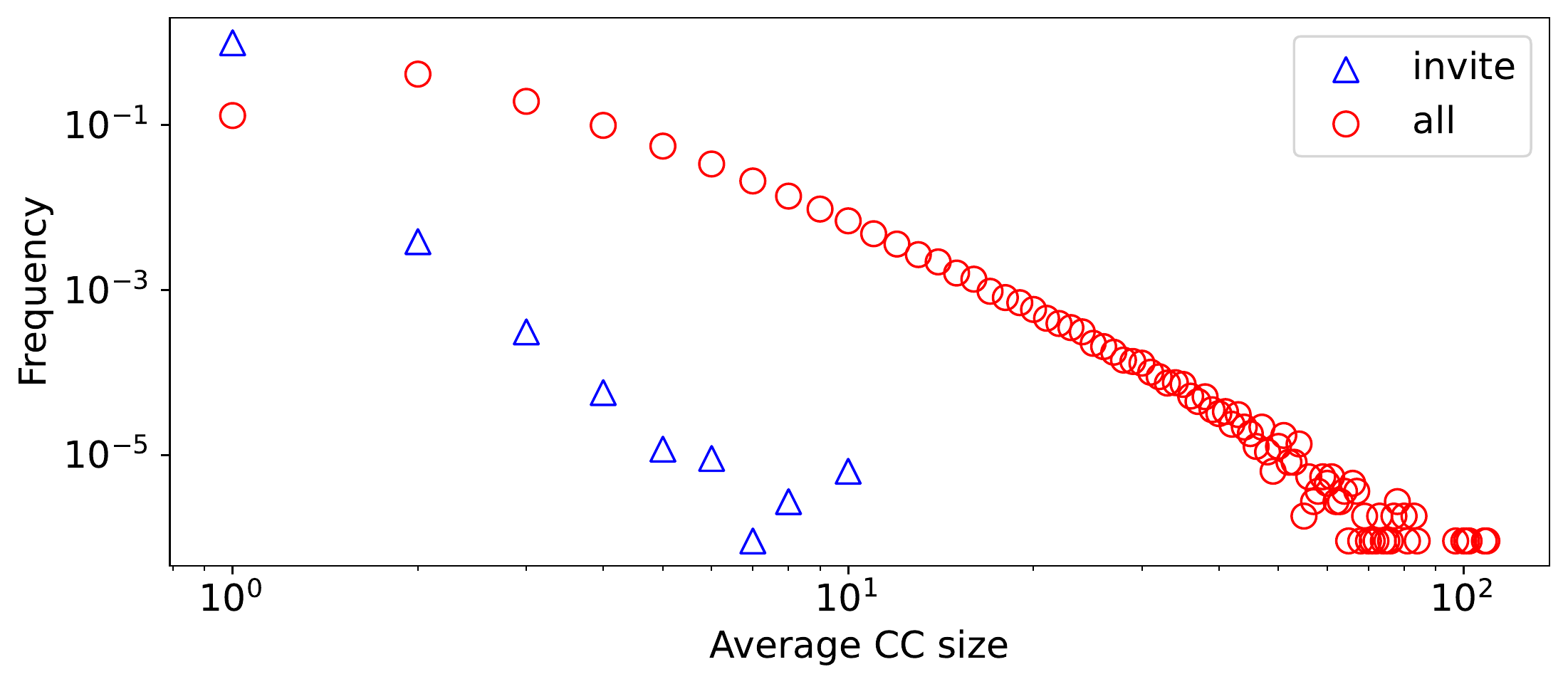}
		\vspace{-2mm}
    \caption{Distribution of averaged $\boldsymbol{\mathsf{CC}}$ size
    in \codmrecall.} \label{fig:distribution}
    \vspace{-2mm}
\end{figure}

The rationales are as follows. First, before the release time of a given event, invitation behaviors of source users are unknown, and hence corresponding ego networks fail to be extracted. Second, even though the ego networks may be constructed by exploiting source behaviors from historical events, however, the invitation behaviors are only triggered by a small number of source users, which renders a much sparser ego network with many singleton \ccs (\ie one inside member)~\cite{ugander2012structural} and makes less group information revealed. 
For example, Figure~\ref{fig:distribution} shows the distributions of averaged \cc size of invited target users on \codmrecall dataset, in which averaged \cc size of a target $v_t$ equals the fraction of the number of all (or inviting) source neighbors over $\#$\cc of $v_t$. Specifically, by using the inviting source nodes, the resulting distribution in blue triangles is highly-skewed, where singletons are over 99\%. In contrast, by using all source neighbors, the results in red circles show an improvement w.r.t the average size, which reveals more group information.

\subsection{Social Standing and Centrality}
Recall in Section~\ref{sec:social-identity} that factors of social standing and centrality in \sit describe the group importance in the whole graph and a given target's cognition, which can be quantified by PageRank and personalized PageRank (PPR), respectively.
Given a graph $\Gset$ with $|\V|=n$ and a restart probability $\alpha$, the PageRank vector of $\Gset$ is denoted as a vector $\boldsymbol{\Pi}\in \mathbb{R}^{1\times n}$, where the $i$-the column value $\boldsymbol{\Pi}[i]={\rho}_i$ is the PageRank value of $v_i$. The PageRank vector $\boldsymbol{\Pi}$ satisfies the following equation
\begin{equation}\label{eq:pr}
    \boldsymbol{\Pi} = \sum_{t=0}^{\infty}{\alpha(1-\alpha)^t\mathbf{s}\cdot\mathbf{P}^t},
\end{equation}
where $\boldsymbol{s}\in \mathbb{R}^{1\times n}$ is the starting vector with the $i$-th column value $\mathbf{s}[i]=1/n$ for each $v_i \in \V$. In Eq.\eqref{eq:pr}, $\mathbf{P}\in \mathbb{R}^{n\times n}$ is the probability transition matrix of $G$ where the value in the $i$-th row and $j$-th column is $\mathbf{P}{[i,j]}={1}/{d_i}$ for $(v_i,v_j)\in \E$ and $\mathbf{P}{[i,j]}=0$ otherwise, where $d_i$ is the number of target neighbors of $v_i$.
Correspondingly, a PPR vector $\mathbf{\Pi}_i\in \mathbb{R}^{1\times n}$ w.r.t the user $v_i$ can be derived by setting the starting vector $\mathbf{s}$ of Eq.\eqref{eq:pr} to an one-hot vector where $\mathbf{s}[i]=1$ for $v_i$ and $\mathbf{s}[i]=0$ otherwise. The value in the $j$-th column $\mathbf{\Pi}_i[j]=\pi_{i,j}$ is the PPR value of $v_j$ w.r.t $v_i$. 

To encapsulate PageRank and PPR into social identity features, we bring the idea from~\cite{everett1999centrality} which proposed to measure the group centrality by taking the average centrality scores of in-group members.
In particular, we define group PageRank (\cpr) and group personalized PageRank (\cppr) as follows.
\begin{definition}[Group PageRank]\label{def:gpr}
Given a graph $\Gset$, a target user $v_t$ and a candidate group $\Cset$, the group PageRank of $\Cset$ is defined as
$$\rho_{\Cset}=\frac{1}{|\Cset|-1}\sum\limits_{v_j\in \Cset\backslash v_t}{\rho}_j.$$
\end{definition}
\begin{definition}[Group Personalized PageRank]\label{def:gppr}
Given a graph $\Gset$, a target user $v_t$ and a candidate group $\Cset$, the group personalized PageRank of $\Cset$ w.r.t $v_t$ is defined as
$$\pi_{t,\Cset}=\frac{1}{|\Cset|-1}\sum\limits_{v_j\in \Cset\backslash v_t}{\pi}_{t,j}.$$
\end{definition}

To understand the intuitions behind \cpr and \cppr, the PageRank $\rho_j$ (resp. PPR $\pi_{t,j}$) in Definition~\ref{def:gpr} (resp. Definition~\ref{def:gppr}) can be interpreted as measuring the global importance and social status of in-group member $v_j$~\cite{song2007identifying,wu2016relational} (resp. relative importance of $v_j$ w.r.t $v_t$~\cite{yang13homogeneous}) by performing the RWR as illustrated in Section~\ref{sec:exist-measure}. More concrete, $\rho_j$ represents the probability that an RWR starting from a randomly-selected node stops at $v_j$. In the meantime, $\pi_{t,j}$ also represents the probability that an RWR stops at $v_j$, but the starting node is the given node $v_t$.
Thus, \cpr and \cppr can indicate the group importance in terms of the probability that a specific RWR terminates in a given group.

Another possible definition for group importance is to replace (personalized) PageRank with the out-degree centrality. However, compared with PageRank, the out-degree centrality $d_i$ of a node $v_i$ only considers the one-hop structural information surrounding $v_i$, which fails to extensively capture the importance of $v_i$ in the whole graph.
Besides the above definitions, \cite{everett1999centrality} also suggests aggregating importance scores of in-group members by summation. Nevertheless, grouping by summation imports the bias from the group scale and yields a misleading high \cpr and \cppr scores for the group, in which many users contain but the importance of each is tiny, as justified in Section~\ref{sec:exp-predict}.

\subsection{Solidarity, Self-stereotyping and In-group Homogeneity}
This part proposes user-group tightness (\ugt) to seamlessly encapsulate solidarity and self-stereotyping factors, and proposes intra-group tightness (\igt) to quantify the in-group homogeneity factor.
More concrete, given a graph $\Gset$ and a node pair $(v_i,v_j)$, we use the tie strength $w_{i,j}$ (resp. the similarity $\delta_{i,j}$) as the backbone for the solidarity factor (resp. self-stereotyping and homogeneity factors). W.l.o.g, the similarity score $\delta_{i,j}\in [0,1]$ is defined as the cosine similarity of the Node2Vec representations~\cite{grover2016node2vec} of $(v_i,v_j)$, which is further normalized to the range of $[0,1]$ by the min-max scaler.
In what follows, we formally define \ugt and \igt as follows.

\begin{definition}[User-Group Tightness]\label{def:ugt}
Given a graph $\Gset$, a target user $v_t$ and a candidate group $\Cset$, the \ugt score of $\Cset$ is defined as
$$\phi_{t,\Cset}=\frac{1}{\sum\limits_{v_j\in \Cset\backslash v_t}w_{t,j}}\sum_{v_j\in\Cset \backslash v_t}{w_{t,j}\cdot\delta_{t,j}}.$$
\end{definition}

\begin{definition}[Intra-Group Tightness]\label{def:igt}
Given a graph $\Gset$ a target user $v_t$ and a candidate group $\Cset$, the \igt score of $\Cset$ is defined as
$$\psi_{\Cset}=\frac{1}{|\Cset|-1}\sum_{v_j\in\Cset \backslash v_t}{\phi_{j,\Cset\backslash v_t}}.$$
\end{definition}

\begin{figure}[t]
\centering
\begin{small}
\subfloat[Local neighborhood]{
    \centering
	\includegraphics[width=0.45\columnwidth]{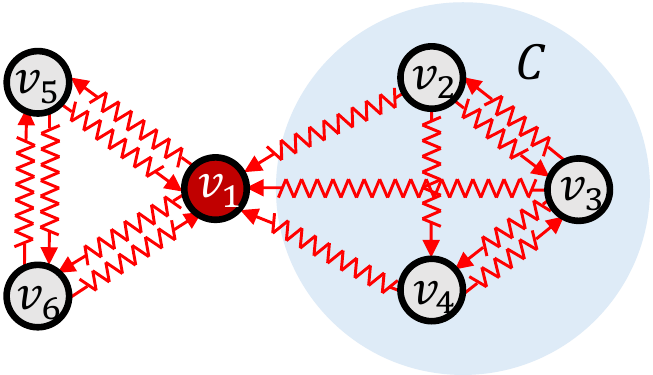}
\vspace{4mm}%
}
\hspace{4mm}
\subfloat[Attractive force]{
\centering
	\includegraphics[width=0.35\columnwidth]{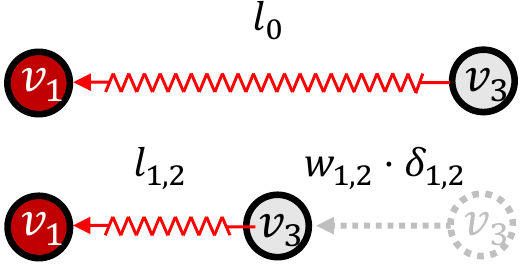}
\vspace{8mm}%
}
\vspace{-2mm}
\end{small}
\caption{Spring-like local neighborhood and attractive force.}
\vspace{-2mm}
\label{fig:spring}
\end{figure}

The intuition behind \ugt (resp. \igt) can be interpreted as the averaged attractive spring-like forces between the target user and other group members (resp. among other group members), if the source neighborhood of the target user is abstracted as a force system. This abstraction is widely adopted in graph drawing~\cite{fruchterman1991graph,gansner2004graph}.
Take Figure~\ref{fig:spring}(a) as an example. The source neighborhood of $v_1$ is represented as a spring network where each edge is assumed as a spring, then the tie strength $w_{i,j}$ can be regarded as the stiffness constant (\ie physical strength) of the spring on $(v_i,v_j)\in \E$.
Let $l_{i,j}\in [0,1]$ be the natural length (when no force is exerted) of a given spring on $(v_i,v_j)$, which is equivalent to the theoretical distance between $v_i$ and $v_j$ on the graph, as clarified in~\cite{gansner2004graph}. Suppose that all springs are set to initial length 1, then the similarity $\delta_{i,j}=1-l_{i,j}$ indicates the displacement of the spring on $(v_i,v_j)$.
According to Hooke's law, as shown in Figure~\ref{fig:spring}(b), $w_{i,j}\cdot\delta_{i,j}$ is equal to the attractive force exerted by $(v_i,v_j)$.
In contrast, group-level competitors in Section~\ref{sec:exist-measure} only consider the tie strength~\cite{song2020you,barbieri2014follow,de2016discriminative} or the in-group connectivity by the edge density~\cite{purohit2014understanding,qiu2016lifecycle}, whereas the homogeneity of group members is ignored.

%% file: exp.tex
\section{Experiments}\label{sec:exp}
In this part, we first elaborate on the experimental settings, and then evaluate the performance of behavior prediction and target recommendation tasks by employing the proposed measures.
All of our experiments are conducted on an in-house cluster consisting of hundreds of machines, each of which runs CentOS, and has 16GB memory and 12 Intel Xeon Processor E5-2670 CPU cores.

\subsection{Experimental Settings} \label{sec:exp-set}
\stitle{Datasets}
We use 3 friendship-enhancing event datasets from Tencent's first person shooter (FPS) and multiuser online battle arena (MOBA) games. 
A given event, whose procedure is explained in Section~\ref{sec:task}, takes the snapshot of $\Gset$ before the release time as the input graph, since $\Gset$ for a particular online game evolves when new users are registered, or friendships are modified. 
We select events \codmrecall and \lgamerecall as datasets to evaluate the performance on target adoption and source invitation behavior predictions, and events \lgamerecall and \xinshi to evaluate the performance on target recommendation.
The statistics of the graph snapshot, source, and target sets for events are summarized in Table~\ref{tab:dataset}. 
All datasets have been anonymized to avoid any leakage of privacy information.

\stitle{Competitors}
We compare the proposed \sit-based measures with 8 representative prior ones as mentioned in Section~\ref{sec:exist-measure}: (i) \textit{individual-level measures}: tie strength (\tie)~\cite{fang2014modeling}, number of common neighbors (\com)~\cite{adamic2003friends}, \ppr~\cite{page1999pagerank}, cosine and euclidean similarity between Node2Vec representations (\ntovcos and \ntoveuc)~\cite{grover2016node2vec}; (ii) \textit{group-level measures}: structural diversity ($\#$\cc)~\cite{ugander2012structural}, user-group tie strength (\gt)~\cite{barbieri2014follow} and group edge density (\gd)~\cite{purohit2014understanding}. All measures are computed based on the network structure as introduced in Table~\ref{tab:dataset}.
Notice in Tencent's MOBA and FPS online gaming platforms that the edge weight between a pair of users is described by the \textit{intimacy} score, which records the number of historical activities/interactions from one to the other, \eg co-playing, gifting, etc.
For a fair comparison, we employ intimacy values to measure the tie strength for all related \sit-based measures and competitors. Furthermore, we employ distributed frameworks in~\cite{lin2019distributed} and ~\cite{lin2021large,lin2020initialization} to compute \ppr and ${\mathsf{N2V}}$-based measures, respectively, and set the final embedding dimension of the latent vector to 200.
To accomplish downstream tasks, we treat our proposal called \sit (including $\#$\cc, \gs, \cpr, \cppr, \ugt, and \igt) or competing measures as input features for~\xgboost as mentioned in Section~\ref{sec:overview}, in which the parameters are set following the original paper~\cite{chen2016xgboost}.

\begin{table}[!t]
\centering
\caption{Dataset statistics ($\boldsymbol{M\!=\!10^6, B\!=\!10^9}$).}\label{tab:dataset}\vspace{-2mm}
\renewcommand{\arraystretch}{1.1}
\begin{small}
\begin{tabular}{|c|c|c|c|c|}
\hline
\textbf{Dataset}     & $\boldsymbol{|\V|}$ & $\boldsymbol{|\E|}$ & $\boldsymbol{|\Sset|}$ & $\boldsymbol{|\Tset|}$ \\ \hline
\codmrecall             &  $77.2M$   &  $1.1B$   &  $33.5M$    & $43.6M$    \\ \hline
\lgamerecall       &  $111.0M$   &  $4.5B$   &  $111.0M$   &  $94.7M$   \\ \hline
\xinshi             &  $130.2M$   &  $6.5B$   &  $120.5M$   &  $99.7M$   \\ \hline
\end{tabular}
\end{small}
\vspace{-2mm}
\end{table}

\input{overall_performance.tex}

\input{ablation_performance.tex}

\subsection{Behavior Prediction}\label{sec:exp-predict}
In this section, we separately train two sets of \xgboost models to evaluate the performance of the aforementioned methods in predicting target adoption and source invitation behaviors.
Notice that user behaviors can be interfered with by the underlying exposure strategies, which will be illustrated in Section~\ref{sec:exp-recommend}.
To eliminate the bias, we focus on randomly exposed source-target pairs, in which we select the pairs with invitation or adoption behaviors as positive data instances and also randomly select an equal number of pairs without behaviors as negative data instances~\cite{zhang2021understanding}. Finally, we obtain 116.3 (resp. 12.5) thousands of data instances for invitation (resp. adoption) prediction on \codmrecall, and 776.4 (resp. 39.1) thousands of instances for invitation (resp. adoption) prediction on \lgamerecall, where 80\% are used for training and 20\% for testing.
To evaluate the effectiveness, we repeat each approach 3 times and report the average result under conventional metrics: Area Under Curve (AUC), accuracy and F1 score.
In the following experiments, we compare our proposal called \sit with each competitor, every single dimension of \sit and their variants in order, followed by analyzing each \sit dimension. Note that we first show the results about target adoptions and then those about source invitations, as the primary goal of \sit to understand how targets are influenced.

\stitle{Overall Performance}
As illustrated in Table~\ref{tab:overall-performance}, our proposed \sit consistently outperforms other competitors in both \codmrecall and \lgamerecall datasets. Specifically, while predicting target adoption behaviors, \sit is 11.8\%, 3.5\%, 2.5\% (resp. 23.2\% 12.6\% 6.7\%) better than the best competitor in terms of AUC, accuracy, and F1 score in \codmrecall (resp. \lgamerecall). For invitation behaviors, \sit is 16.2\%, 12.3\%, 16.7\% (resp. 12.0\%, 8.1\%, and 8.7\%) better than the best competitor in terms of AUC, accuracy, and F1 score in \codmrecall (resp. \lgamerecall).
The above results reflect that involving group information by \sit is necessary for understanding user behaviors in friendship-enhancing events. 
Regarding competitors, \tie is the best-performing competitor in terms of AUC and accuracy scores during the adoption prediction, which indicates that source-target pairs with more historical interactions (\ie intimacy) are more likely to interact again. Correspondingly, AUC and accuracy scores of \gt and \gd are also comparable due to the usage of intimacy values. Furthermore, the F1 score of other single-level measures (\ppr, \com or \ntovcos) is the second best on both datasets w.r.t. adoption and invitation predictions. Regarding the structural diversity ($\#$\cc), we evaluate its performance in the ablation study, as it is a dimension of \sit.

\stitle{Ablation study}
As illustrated in Table~\ref{tab:ablation-performance}, we find that each dimension \ie $\#$\cc, \gs, \cpr, \cppr, \ugt, or \igt, is less effective than \sit. For both behaviors and datasets, \ugt or \cpr performs as the best competitor in all evaluation metrics, compared with both other \sit-based dimensions and all competitors in Table~\ref{tab:overall-performance}. Regarding \igt, it is less effective than \ugt and \cpr, but is still comparable to the existing measures in Table~\ref{tab:overall-performance}. For instance, the \igt score is 4.3\%, 5.6\% better than the group-level competitor \gd in terms of AUC and accuracy for adoption behaviors of \lgamerecall, as \igt additionally involves the similarities among in-group members.
Notice that both $\#$\cc and \gs scores are analyzed in the work of structural diversity~\cite{ugander2012structural} and can be treated as competitors of other \sit-based measures. In particular, we assemble both measures into \xgboost and denote the model instance as $\#$\cc-\gs, whose scores can extensively outperform those of each single measure $\#$\cc and \gs as reported in Table~\ref{tab:ablation-performance}. However, $\#$\cc-\gs are still worse than some single dimensions (\ie \cpr or \ugt) and the proposed \sit. 
Regarding the reasons for the performance of each dimension in \sit, we leave them in the analysis parts.

\input{codm_plot.tex}

\stitle{Factor formulation}
As mentioned in Section~\ref{sec:metric}, other possible formulations of proposed measures exist. Thus, except for the averaged aggregation operation and cosine similarity, we also evaluate the performance of variant measures with summation operation and euclidean similarity (denoted as sum and euc, respectively) in Table~\ref{tab:ablation-performance}. To derive the euclidean similarity, we first normalize the euclidean distance between each node pair by the min-max scaler as the cosine similarity does, and then take the one minus normalized distance as the corresponding similarity.
In particular, employing the summation operation is equivalent to hybrid the original measure with the factor of \gs, which may dampen the original measure (\ie \cpr, \cppr, \ugt or \igt) because of the weak correlation between \gs and user behaviors, which will be shown in Figures~\ref{fig:codm-adopt}-\ref{fig:codm-invite}. For example, the result quality of the summation of \cpr is consistently worse than that of \cpr on both behaviors and datasets.
As for different similarity measures w.r.t node representations, the effectiveness of utilizing cosine and euclidean similarity is comparable. In particular, the F1 scores of euclidean-based \ugt and \igt are usually better than those of cosine-based. Meanwhile, negative results are reported regarding AUC and accuracy scores.

\stitle{Conversion probability conditioned on \sitb dimensions}
We investigate the relation between user behaviors and the aforementioned \sit-based measures: $\#$\cc, \gs, \cpr, \cppr, \ugt, and \igt. Due to space limitations, we only report the analysis results on \codmrecall, and the rest datasets derive similar results.
Following the setting of~\cite{fang2014modeling}, we employ the density-based discretization to convert the scores of \sit-based measures into five different levels, where a higher level indicates a larger score.
Moreover, we use \textit{conversion probability}~\cite{fang2014modeling, ugander2012structural} to evaluate how the user acts. In particular, given a level of certain \sit measure, the adoption (resp. invitation) conversion probability is the fraction of source-target pairs existing adoption (resp. invitation) behaviors over all pairs in this level.

We report the adoption and invitation conversion probabilities w.r.t each measure in Figure~\ref{fig:codm-adopt} and Figure~\ref{fig:codm-invite}, respectively. We can find an evident correlation between each behavior and each proposed measure except for \gs.
In terms of $\#$\cc, contradicting the well-accepted insight of structural diversity~\cite{ugander2012structural,su2020experimental,fang2014modeling}, Figures~\ref{fig:codm-adopt}(a) and ~\ref{fig:codm-invite}(a) show that the target user with less number of candidate groups (\ie a less diverse neighborhood) is more likely to be invited and further adopt this invitation. Recently, this phenomenon also emerged from other user behaviors, especially in Tencent instant messaging platform~\cite{qiu2016lifecycle,zhang2021understanding}, but the underlying reason is still unclear. In contrast, \sit can explain this as that a greater multi-membership makes the social identification phase of a target challenging due to the potential role conflict~\cite{marks1977multiple}.
Regarding the effect of group size, we can find in Figures~\ref{fig:codm-adopt}(b) and ~\ref{fig:codm-invite}(b) that the engagement willingness of users will be facilitated as the group scale increases, but subsequently hurdled if still going up. Notice that the source-target pair inside a medium-size group (\ie level 3) is more likely to send an invitation and respond.

Recall in Section~\ref{sec:metric} that \cpr and \cppr measure the factors of social standing and centrality from the global and personalized view, respectively. The results in Figures~\ref{fig:codm-adopt}(d) and \ref{fig:codm-invite}(d) show that the target user is more likely to choose a relatively important group, which confirms the claim of the centrality factor in \sit. However, Figures~\ref{fig:codm-adopt}(c) and \ref{fig:codm-invite}(c) illustrate opposing results w.r.t \cpr. The reason is that, by definition, a user with a high PageRank value tends to have more neighbors, which consequently means each target friend has less probability of being exposed by the random algorithm.

To evaluate the tie strength and similarity parts in \ugt, we separate them as \ugt-$w$ and \ugt-$\delta$ respectively, in which \ugt-$w$ measures the averaged intimacy values and \ugt-$\delta$ measures the averaged cosine similarity between the Node2Vec representations. The same operation is also performed for \igt. As illustrated in Figures~\ref{fig:codm-adopt}(e) and \ref{fig:codm-invite}(e), the behavior conversion probabilities increase as the levels of \ugt-$w$ and \ugt-$\delta$ go up, which match the factors of solidarity and self-stereotyping in \sit. The same tendency for \igt is shown in Figures~\ref{fig:codm-adopt}(f) and \ref{fig:codm-invite}(f). However, the correlation regarding \igt is relatively weak as it fails to provide intra-group information for the group with only one member.

\input{feature_importance.tex}

\stitle{Importance of \sitb dimensions}
\xgboost supports measuring the importance of each input feature, where the importance indicates the frequency that a feature is leveraged to split the data across all trees.
Table~\ref{tab:adopt-importance} and Table \ref{tab:invite-importance} report the feature importance of each proposed \sit-based measure w.r.t adoption and invitation behaviors, respectively. In particular, \cpr, \cppr, and \ugt have the top-3 highest importance w.r.t both behaviors, which indicates that social standing, centrality, solidarity, and self-stereotyping are more critical \sit factors. As for \gs and \igt, they perform as the bottom-2 important features. This is because they are less correlated with user behaviors, as illustrated in Figures~\ref{fig:codm-adopt}-\ref{fig:codm-invite}.
In contrast, $\#$\cc is also less important even though it has quite strong correlation to user behaviors. This is due to that $\#$\cc is a stand-alone measure for the target user and hence incurs indistinguishable values for source-target pairs with the same target user.  
To summarize, the aforementioned results of the correlation and importance analysis can explain each \sit dimension's performance in Table~\ref{tab:ablation-performance}.

\subsection{Target Recommendation}\label{sec:exp-recommend}

\stitle{Deployment setups}
Recall in Section~\ref{sec:task} that
a source user can only select the target users from a feed window of limited size $k$ during the second step of a friendship-enhancing event. 
Therefore, judiciously exposing a subset of target friends for the source $v_i$ is pivotal to the event's performance. Motivated by this, the target recommendation task (Problem~\ref{prob:recommend}) is to find $k$ target users w.r.t the source to boost the overall engagement of sources and targets.
Here, we inherit the \xgboost model, which takes the aforementioned individual measures (\ie \tie (measured by intimacy), \com, \ppr, and \ntoveuc) or \sit measures as the input and is trained based on the historical events. We sort the predicted value $\hat{y}_{i,j}$ w.r.t each source $v_i$ in a descending order and select the top-$k$ target nodes to recommend.
To evaluate the proposed \sit measures, we conduct the online A/B testing that randomly assigns a fraction of live traffic to \xgboost models with individual-level measures as treatment groups. Initially, each measure is computed based on the graph instance ahead of the event (see Table~\ref{tab:dataset}). Afterward, each measure is updated daily by using the latest graph snapshot.

\begin{table}[!t]
\centering
\renewcommand{\arraystretch}{1.2}
\caption{Online performance in \lgamerecall.}\vspace{-2mm}
\begin{small}
\begin{tabular}{|c|c|c|c|c|c|}
\hline
\textbf{Measure} & \tie & \com & \ppr   & \ntoveuc & \sit                            \\ \hline
\textbf{E2E rate}    & 0.1018   & 0.0958       & \underline{0.1066} & 0.0739        & \cellcolor{blue!25} 0.1431 \\ \hline
\end{tabular}
\end{small}\label{tab:online-lgamerecall}
\vspace{-2mm}
\end{table}

\begin{table}[!t]
\centering
\renewcommand{\arraystretch}{1.2}
\caption{Online performance in \xinshi.}\vspace{-2mm}
\begin{small}
\begin{tabular}{|c|c|c|c|}
\hline
\textbf{Measure} & \tie & \ppr    & \sit      \\ \hline
\textbf{E2E rate}    & 0.1152   & \underline{0.1218} & \cellcolor{blue!25} 0.1384 \\ \hline
\end{tabular}
\end{small}\label{tab:online-xinshi}
\vspace{-2mm}
\end{table}

\stitle{Overall performance}
We evaluate the effectiveness of different online random trials by the end-to-end (E2E) rate, which considers the overall engagement of sources and targets. In particular, E2E rate equals the fraction of target friends adopting the invitations over the source users seeing the event. The higher E2E rate indicates better quality.
As illustrated in Table~\ref{tab:online-lgamerecall}, the proposed solution \sit has the highest E2E rate on \lgamerecall.
Specifically, \sit is 34.2\% better than the best-performed treatment approach \ppr.
Furthermore, we can find that \ppr and \tie can outperform the other two competitors. Motivated by this, we conduct another online trial on \xinshi by comparing \sit with the best two measures \ppr and \tie. Specifically, Table~\ref{tab:online-xinshi} reports that \sit is 13.6\% better than the best-performed treatment approach \ppr.

%% file: overall_performance.tex
\begin{table*}[!ht]
\centering
\renewcommand{\arraystretch}{1.2}
\caption{Comparison of existing measures and proposed \sitb measure (the best is colored and the second best is underlined).}\vspace{-2mm}
\begin{small}
\begin{tabular}{|c|cccccc|cccccc|}
\hline
\multirow{3}{*}{\textbf{Measure}} & \multicolumn{6}{c|}{\textbf{Adoption}}                                                    & \multicolumn{6}{c|}{\textbf{Invitation}}                                                  \\ \cline{2-13} 
                         & \multicolumn{3}{c|}{\bf{\codmrecall}}                         & \multicolumn{3}{c|}{\bf{\lgamerecall}}   & \multicolumn{3}{c|}{\bf{\codmrecall}}                         & \multicolumn{3}{c|}{\bf{\lgamerecall}}   \\ \cline{2-13} 
                         & \textbf{AUC}    & \textbf{Accuracy} & \multicolumn{1}{c|}{\textbf{F1 score}} & \textbf{AUC}    & \textbf{Accuracy} & \textbf{F1 score} & \textbf{AUC}    & \textbf{Accuracy} & \multicolumn{1}{c|}{\textbf{F1 score}} & \textbf{AUC}    & \textbf{Accuracy} & \textbf{F1 score} \\ \hline
\tie                      & \underline{0.7154} & \underline{0.6965}   & \multicolumn{1}{c|}{0.6554}   & \underline{0.6017} & \underline{0.6021}   & 0.3958   & 0.6072 & \underline{0.5985}   & \multicolumn{1}{c|}{0.4607}   & 0.5361 & 0.5353   & 0.2200   \\
\com                      & 0.5488 & 0.5538   & \multicolumn{1}{c|}{0.5615}   & 0.5667 &	0.5576 &	\underline{0.6219}    & 0.5456 & 0.5323   & \multicolumn{1}{c|}{0.4674}   & 0.5332 & 0.5281   & \underline{0.5565}   \\
\ppr                      & 0.6565 & 0.6036   & \multicolumn{1}{c|}{0.5596}   & 0.5562 & 0.5388   & 0.4447   & \underline{0.6289} & 0.5972   & \multicolumn{1}{c|}{\underline{0.5786}}   & \underline{0.5846} & 0.5589   & 0.5467   \\
\ntovcos                      & 0.6976 & 0.6610   & \multicolumn{1}{c|}{\underline{0.7171}}   & 0.5808 & 0.5626   & 0.5420   & 0.5608 & 0.5537   & \multicolumn{1}{c|}{0.5683}   & 0.5770 & \underline{0.5630}   & 0.5426   \\
\ntoveuc                      & 0.7076 & 0.6652   & \multicolumn{1}{c|}{0.7091}   & 0.5664 & 0.5566   & 0.5390   & 0.5679 & 0.5588   & \multicolumn{1}{c|}{0.5628}   & 0.5739 & 0.5585   & 0.5375   \\ \hline
\gt                  & 0.6985 & 0.6572   & \multicolumn{1}{c|}{0.6004}   & 0.6295 & 0.5959   & 0.5777   & 0.5738 & 0.5652   & \multicolumn{1}{c|}{0.3988}   & 0.5397 & 0.5297   & 0.4875   \\
\gd                   & 0.6077 & 0.5736   & \multicolumn{1}{c|}{0.5269}   & 0.6039 & 0.5728   & 0.5908   & 0.5811 & 0.5507   & \multicolumn{1}{c|}{0.4490}   & 0.5674 & 0.5508   & 0.4991   \\ \hline
\sit                      & \cellcolor{blue!25}0.7995 & \cellcolor{blue!25}0.7206   & \multicolumn{1}{c|}{\cellcolor{blue!25}0.7350}   & \cellcolor{blue!25}0.7410 & \cellcolor{blue!25}0.6780   & \cellcolor{blue!25}0.6638   & \cellcolor{blue!25}\cellcolor{blue!25}0.7307 & \cellcolor{blue!25}0.6719   & \multicolumn{1}{c|}{\cellcolor{blue!25}0.6754}   & \cellcolor{blue!25}0.6550 & \cellcolor{blue!25}0.6086   & \cellcolor{blue!25}0.6047   \\ \hline
\end{tabular}
\end{small}\label{tab:overall-performance}
\vspace{0mm}
\end{table*}

%% file: ablation_performance.tex
\begin{table*}[!t]
\centering
\renewcommand{\arraystretch}{1.2}
\caption{Comparison of different \sitb variants (the best is colored and the second best is underlined).}\vspace{-2mm}
\begin{small}
\begin{tabular}{|c|cccccc|cccccc|}
\hline
\multirow{3}{*}{\textbf{Measure}} & \multicolumn{6}{c|}{\textbf{Adoption}}                                                    & \multicolumn{6}{c|}{\textbf{Invitation}}                                                  \\ \cline{2-13} 
                         & \multicolumn{3}{c|}{\bf{\codmrecall}}                         & \multicolumn{3}{c|}{\bf{\lgamerecall}}   & \multicolumn{3}{c|}{\bf{\codmrecall}}                         & \multicolumn{3}{c|}{\bf{\lgamerecall}}   \\ \cline{2-13} 
                         & \textbf{AUC}    & \textbf{Accuracy} & \multicolumn{1}{c|}{\textbf{F1 score}} & \textbf{AUC}    & \textbf{Accuracy} & \textbf{F1 score} & \textbf{AUC}    & \textbf{Accuracy} & \multicolumn{1}{c|}{\textbf{F1 score}} & \textbf{AUC}    & \textbf{Accuracy} & \textbf{F1 score} \\ \hline
$\#$\cc                       & 0.6153 &	0.5897 	 & \multicolumn{1}{c|}{0.5378}   & 0.5452 & 0.5288   & 0.4136   & 0.6091 &	0.5820  & \multicolumn{1}{c|}{0.5392}   & 0.5790 & 0.5551   & 0.5662   \\
\gs                       & 0.5866 & 0.5703   & \multicolumn{1}{c|}{0.5323}   & 0.5860 & 0.5720   & 0.6182   & 0.5669 & 0.5448   & \multicolumn{1}{c|}{0.4827}   & 0.5655 & 0.5510   & 0.5266   \\ \hline
\cpr                      & 0.6509 & 0.6071   & \multicolumn{1}{c|}{0.6150}   & \underline{0.6567} & \underline{0.6149}   & \underline{0.6403}   & \underline{0.6817} & \underline{0.6314}   & \multicolumn{1}{c|}{\underline{0.6325}}   & \underline{0.6382} & \underline{0.5978}   & \underline{0.6044}   \\
-sum                  & 0.5750 & 0.5538   & \multicolumn{1}{c|}{0.5702}   & 0.5651 & 0.5469   & 0.5798   & 0.6074 & 0.5699   & \multicolumn{1}{c|}{0.4978}   & 0.5783 & 0.5531   & 0.5545   \\ \hline
\cppr                     & 0.6337 & 0.5990   & \multicolumn{1}{c|}{0.5630}   & 0.5641 & 0.5405   & 0.4433   & 0.6258 & 0.5922   & \multicolumn{1}{c|}{0.5747}   & 0.5895 & 0.5603   & 0.5902   \\
-sum                 & 0.6230 & 0.5727   & \multicolumn{1}{c|}{0.5749}   & 0.5874 & 0.5671   & 0.6217   & 0.5771 & 0.5472   & \multicolumn{1}{c|}{0.5634}   & 0.5683 & 0.5493   & 0.5545   \\ \hline
\ugt                      & \underline{0.7427} & \underline{0.6965}   & \multicolumn{1}{c|}{0.7249}   & 0.5634 & 0.5482   & 0.5061   & 0.5969 & 0.5772   & \multicolumn{1}{c|}{0.5793}   & 0.5684 & 0.5525   & 0.5345   \\
-euc                  & 0.7233 & 0.6857   & \multicolumn{1}{c|}{\underline{0.7266}}   & 0.5492 & 0.5345   & 0.5199   & 0.5844 & 0.5643   & \multicolumn{1}{c|}{0.5761}   & 0.5650 & 0.5505   & 0.5402   \\
-sum                  & 0.7374 & 0.6621   & \multicolumn{1}{c|}{0.6145}   & 0.6380 & 0.6035   & 0.5793   & 0.5920 & 0.5695   & \multicolumn{1}{c|}{0.4477}   & 0.5574 & 0.5378   & 0.5695   \\ \hline
\igt                      & 0.5622 & 0.5425   & \multicolumn{1}{c|}{0.4388}   & 0.6296 & 0.6047   & 0.5487   & 0.5372 & 0.5236   & \multicolumn{1}{c|}{0.3694}   & 0.5642 & 0.5496   & 0.4616   \\
-euc                  & 0.5645 & 0.5650   & \multicolumn{1}{c|}{0.4670}   & 0.5733 & 0.5569   & 0.6265   & 0.5344 & 0.5239   & \multicolumn{1}{c|}{0.3617}   & 0.5681 & 0.5521   & 0.4660   \\
-sum                  & 0.5638 & 0.5560   & \multicolumn{1}{c|}{0.5008}   & 0.5915 & 0.5708   & 0.6058   & 0.5620 & 0.5373   & \multicolumn{1}{c|}{0.4605}   & 0.5705 & 0.5531   & 0.5170   \\ \hline
$\#$\cc-\gs                      & 0.6698 & 0.6180   & \multicolumn{1}{c|}{0.6038}   & 0.6101 & 0.5857   & 0.6252   & 0.6306 & 0.5978   & \multicolumn{1}{c|}{0.5808}   & 0.5972 & 0.5665   & 0.5550   \\
\sit                      & \cellcolor{blue!25}0.7995 & \cellcolor{blue!25}0.7206   & \multicolumn{1}{c|}{\cellcolor{blue!25}0.7350}   & \cellcolor{blue!25}0.7410 & \cellcolor{blue!25}0.6780   & \cellcolor{blue!25}0.6638   & \cellcolor{blue!25}0.7307 & \cellcolor{blue!25}0.6719   & \multicolumn{1}{c|}{\cellcolor{blue!25}0.6754}   & \cellcolor{blue!25}0.6550 & \cellcolor{blue!25}0.6086   & \cellcolor{blue!25}0.6047   \\ \hline
\end{tabular}
\end{small}\label{tab:ablation-performance}
\vspace{0mm}
\end{table*}

%% file: codm_plot.tex
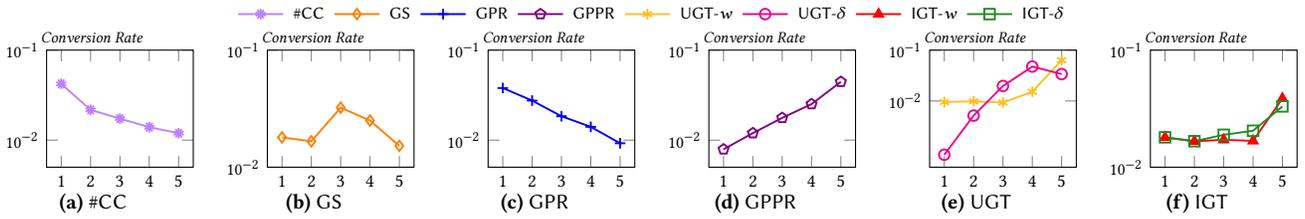
\begin{figure*}[!t]
\centering
\begin{small}
\begin{tikzpicture}
    \begin{customlegend}[legend columns=8,
        legend entries={$\#$\cc,\gs, \cpr, \cppr, \ugt-$w$,\ugt-$\delta$, \igt-$w$,\igt-$\delta$},
        legend style={at={(0.45,1.15)},anchor=north,draw=none,font=\scriptsize,column sep=0.05cm}]
        \addlegendimage{line width=0.25mm,mark=10-pointed star,color=RYB1}
        \addlegendimage{line width=0.25mm,color=orange,mark=diamond}
        \addlegendimage{line width=0.25mm,color=blue,mark=+}
        \addlegendimage{line width=0.25mm,color=violet,mark=pentagon}
        \addlegendimage{line width=0.25mm,color=RYB2,mark=asterisk}
        \addlegendimage{line width=0.25mm,color=magenta,mark=o}
        \addlegendimage{line width=0.25mm,color=red,mark=triangle*}
        \addlegendimage{line width=0.25mm,color=forestgreen,mark=square}
    \end{customlegend}
\end{tikzpicture}
\\[-\lineskip]
\vspace{-1mm}
\hspace{-2mm}
\subfloat[$\#$\cc]{
\begin{tikzpicture}[scale=1]
    \begin{axis}[
        height=\columnwidth/2.7,
        width=\columnwidth/2.4,
        ylabel={\em Conversion Rate},
        xmin=0.5, xmax=5.5,
        ymin=0.005, ymax=0.1,
        xtick={1,2,3,4,5},
        xticklabel style = {font=\footnotesize},
        xticklabels={1,2,3,4,5},
        ytick={0.01,0.1},
        yticklabels={$10^{-2}$,$10^{-1}$},
        ymode=log,
        log basis y={10},
        every axis y label/.style={at={(current axis.north west)},right=6mm,above=0mm},
        label style={font=\scriptsize},
        tick label style={font=\scriptsize},
    ]
    \addplot[line width=0.25mm,mark=10-pointed star,color=RYB1] 
        plot coordinates {
(1,	0.042241161)
(2,	0.021686136)
(3,	0.01733687)
(4,	0.013924372)
(5,	0.011935176)
        };    
    \end{axis}
\end{tikzpicture}\vspace{1mm}%
}
\hspace{2mm}
\subfloat[\gs]{
\begin{tikzpicture}[scale=1]
    \begin{axis}[
        height=\columnwidth/2.7,
        width=\columnwidth/2.4,
        ylabel={\em Conversion Rate},
        xmin=0.5, xmax=5.5,
        ymin=0.01, ymax=0.1,
        xtick={1,2,3,4,5},
        xticklabel style = {font=\footnotesize},
        xticklabels={1,2,3,4,5},
        ytick={0.01,0.1},
        yticklabels={$10^{-2}$,$10^{-1}$},
        ymode=log,
        log basis y={10},
        every axis y label/.style={at={(current axis.north west)},right=6mm,above=0mm},
        label style={font=\scriptsize},
        tick label style={font=\scriptsize},
    ]
        \addplot[line width=0.25mm,color=orange,mark=diamond] 
        plot coordinates {
(1,	0.017980771)
(2,	0.016635139)
(3,	0.03227505)
(4,	0.025020965)
(5,	0.015211498)
        }; 
    \end{axis}
\end{tikzpicture}\vspace{1mm}%
}
\hspace{2mm}
\subfloat[\cpr]{
\begin{tikzpicture}[scale=1]
    \begin{axis}[
        height=\columnwidth/2.7,
        width=\columnwidth/2.4,
        ylabel={\em Conversion Rate},
        xmin=0.5, xmax=5.5,
        ymin=0.005, ymax=0.1,
        xtick={1,2,3,4,5},
        xticklabel style = {font=\footnotesize},
        xticklabels={1,2,3,4,5},
        ytick={0.01,0.1},
        yticklabels={$10^{-2}$,$10^{-1}$},
        ymode=log,
        log basis y={10},
        every axis y label/.style={at={(current axis.north west)},right=6mm,above=0mm},
        label style={font=\scriptsize},
        tick label style={font=\scriptsize},
    ]
        \addplot[line width=0.25mm,color=blue,mark=+] 
        plot coordinates {
(1,	0.037931236)
(2,	0.027497709)
(3,	0.018428956)
(4,	0.014041383)
(5,	0.009224409)
        };  
    \end{axis}
\end{tikzpicture}\vspace{1mm}%
}
\hspace{2mm}
\subfloat[\cppr]{
\begin{tikzpicture}[scale=1]
    \begin{axis}[
        height=\columnwidth/2.7,
        width=\columnwidth/2.4,
        ylabel={\em Conversion Rate},
        xmin=0.5, xmax=5.5,
        ymin=0.005, ymax=0.1,
        xtick={1,2,3,4,5},
        xticklabel style = {font=\footnotesize},
        xticklabels={1,2,3,4,5},
        ytick={0.01,0.1},
        yticklabels={$10^{-2}$,$10^{-1}$},
        ymode=log,
        log basis y={10},
        every axis y label/.style={at={(current axis.north west)},right=6mm,above=0mm},
        label style={font=\scriptsize},
        tick label style={font=\scriptsize},
    ]
        \addplot[line width=0.25mm,color=violet,mark=pentagon] 
        plot coordinates {
(1,	0.007898278)
(2,	0.011954678)
(3,	0.017668396)
(4,	0.025137976)
(5,	0.04446438)
        };  
    \end{axis}
\end{tikzpicture}\vspace{1mm}%
}
\hspace{2mm}
\subfloat[\ugt]{
\begin{tikzpicture}[scale=1]
    \begin{axis}[
        height=\columnwidth/2.7,
        width=\columnwidth/2.4,
        ylabel={\em Conversion Rate},
        xmin=0.5, xmax=5.5,
        ymin=0.0005, ymax=0.1,
        xtick={1,2,3,4,5},
        xticklabel style = {font=\footnotesize},
        xticklabels={1,2,3,4,5},
        ytick={0.01,0.1},
        yticklabels={$10^{-2}$,$10^{-1}$},
        ymode=log,
        log basis y={10},
        every axis y label/.style={at={(current axis.north west)},right=6mm,above=0mm},
        label style={font=\scriptsize},
        tick label style={font=\scriptsize},
    ]
        \addplot[line width=0.25mm,color=RYB2,mark=asterisk] 
        plot coordinates {
(1,	0.009555941)
(2,	0.00984847)
(3,	0.009321736)
(4,	0.015191996)
(5,	0.063205726)
        };  
        \addplot[line width=0.25mm,color=magenta,mark=o] 
        plot coordinates {
(1,	0.000878)
(2,	0.005168009)
(3,	0.019735559)
(4,	0.047526181)
(5,	0.033816331)
        }; 
    \end{axis}
\end{tikzpicture}\vspace{1mm}%
}
\hspace{2mm}
\subfloat[\igt]{
\begin{tikzpicture}[scale=1]
    \begin{axis}[
        height=\columnwidth/2.7,
        width=\columnwidth/2.4,
        ylabel={\em Conversion Rate},
        xmin=0.5, xmax=5.5,
        ymin=0.01, ymax=0.1,
        xtick={1,2,3,4,5},
        xticklabel style = {font=\footnotesize},
        xticklabels={1,2,3,4,5},
        ytick={0.01,0.1},
        yticklabels={$10^{-2}$,$10^{-1}$},
        ymode=log,
        log basis y={10},
        every axis y label/.style={at={(current axis.north west)},right=6mm,above=0mm},
        label style={font=\scriptsize},
        tick label style={font=\scriptsize},
    ]
        \addplot[line width=0.25mm,color=red,mark=triangle*] 
        plot coordinates {
(1,	0.017980771)
(2,	0.016635139)
(3,	0.017141854)
(4,	0.016771652)
(5,	0.038594302)
        };  
        \addplot[line width=0.25mm,color=forestgreen,mark=square] 
        plot coordinates {
(1,	0.018000273)
(2,	0.016674142)
(3,	0.018916494)
(4,	0.020477017)
(5,	0.033055756)
        }; 
    \end{axis}
\end{tikzpicture}\vspace{1mm}%
}
\vspace{-3mm}
\end{small}
\caption{Conversion probability of adoption behaviors conditioned on each \sitb-based measure in \codmrecall.} \label{fig:codm-adopt}
\end{figure*}

\begin{figure*}[t]
\centering
\begin{small}
\hspace{-2mm}
\subfloat[$\#$\cc]{
\begin{tikzpicture}[scale=1]
    \begin{axis}[
        height=\columnwidth/2.7,
        width=\columnwidth/2.4,
        ylabel={\em Conversion Rate},
        xmin=0.5, xmax=5.5,
        ymin=0.05, ymax=1,
        xtick={1,2,3,4,5},
        xticklabel style = {font=\footnotesize},
        xticklabels={1,2,3,4,5},
        ytick={0.1,1},
        yticklabels={$10^{-1}$,$10^{0}$},
        ymode=log,
        log basis y={10},
        every axis y label/.style={at={(current axis.north west)},right=6mm,above=0mm},
        label style={font=\scriptsize},
        tick label style={font=\scriptsize},
    ]
    \addplot[line width=0.25mm,mark=10-pointed star,color=RYB1] 
        plot coordinates {
(1,	0.317744798)
(2,	0.23492014)
(3,	0.19544444)
(4,	0.166585409)
(5,	0.141564444)
        };    
    \end{axis}
\end{tikzpicture}\vspace{1mm}%
}
\hspace{2mm}
\subfloat[\gs]{
\begin{tikzpicture}[scale=1]
    \begin{axis}[
        height=\columnwidth/2.7,
        width=\columnwidth/2.4,
        ylabel={\em Conversion Rate},
        xmin=0.5, xmax=5.5,
        ymin=0.1, ymax=1,
        xtick={1,2,3,4,5},
        xticklabel style = {font=\footnotesize},
        xticklabels={1,2,3,4,5},
        ytick={0.1,1},
        yticklabels={$10^{-1}$,$10^{0}$},
        ymode=log,
        log basis y={10},
        every axis y label/.style={at={(current axis.north west)},right=6mm,above=0mm},
        label style={font=\scriptsize},
        tick label style={font=\scriptsize},
    ]
        \addplot[line width=0.25mm,color=orange,mark=diamond] 
        plot coordinates {
(1,	0.20771496)
(2,	0.211420325)
(3,	0.260774601)
(4,	0.225169179)
(5,	0.151178891)
        }; 
    \end{axis}
\end{tikzpicture}\vspace{1mm}%
}
\hspace{2mm}
\subfloat[\cpr]{
\begin{tikzpicture}[scale=1]
    \begin{axis}[
        height=\columnwidth/2.7,
        width=\columnwidth/2.4,
        ylabel={\em Conversion Rate},
        xmin=0.5, xmax=5.5,
        ymin=0.05, ymax=1,
        xtick={1,2,3,4,5},
        xticklabel style = {font=\footnotesize},
        xticklabels={1,2,3,4,5},
        ytick={0.1,1},
        yticklabels={$10^{-1}$,$10^{0}$},
        ymode=log,
        log basis y={10},
        every axis y label/.style={at={(current axis.north west)},right=6mm,above=0mm},
        label style={font=\scriptsize},
        tick label style={font=\scriptsize},
    ]
        \addplot[line width=0.25mm,color=blue,mark=+] 
        plot coordinates {
(1,	0.401427541)
(2,	0.246250756)
(3,	0.181461835)
(4,	0.131072411)
(5,	0.096046961)
        };  
    \end{axis}
\end{tikzpicture}\vspace{1mm}%
}
\hspace{2mm}
\subfloat[\cppr]{
\begin{tikzpicture}[scale=1]
    \begin{axis}[
        height=\columnwidth/2.7,
        width=\columnwidth/2.4,
        ylabel={\em Conversion Rate},
        xmin=0.5, xmax=5.5,
        ymin=0.05, ymax=1,
        xtick={1,2,3,4,5},
        xticklabel style = {font=\footnotesize},
        xticklabels={1,2,3,4,5},
        ytick={0.1,1},
        yticklabels={$10^{-1}$,$10^{0}$},
        ymode=log,
        log basis y={10},
        every axis y label/.style={at={(current axis.north west)},right=6mm,above=0mm},
        label style={font=\scriptsize},
        tick label style={font=\scriptsize},
    ]
        \addplot[line width=0.25mm,color=violet,mark=pentagon] 
        plot coordinates {
(1,	0.083702245)
(2,	0.117772101)
(3,	0.173778228)
(4,	0.247011331)
(5,	0.433995749)
        };  
    \end{axis}
\end{tikzpicture}\vspace{1mm}%
}
\hspace{2mm}
\subfloat[\ugt]{
\begin{tikzpicture}[scale=1]
    \begin{axis}[
        height=\columnwidth/2.7,
        width=\columnwidth/2.4,
        ylabel={\em Conversion Rate},
        xmin=0.5, xmax=5.5,
        ymin=0.1, ymax=1,
        xtick={1,2,3,4,5},
        xticklabel style = {font=\footnotesize},
        xticklabels={1,2,3,4,5},
        ytick={0.1,1},
        yticklabels={$10^{-1}$,$10^{0}$},
        ymode=log,
        log basis y={10},
        every axis y label/.style={at={(current axis.north west)},right=6mm,above=0mm},
        label style={font=\scriptsize},
        tick label style={font=\scriptsize},
    ]
        \addplot[line width=0.25mm,color=RYB2,mark=asterisk] 
        plot coordinates {
(1,	0.171129356)
(2,	0.169588704)
(3,	0.174148758)
(4,	0.176375373)
(5,	0.365017454)
        };  
        \addplot[line width=0.25mm,color=magenta,mark=o] 
        plot coordinates {
(1,	0.175770813)
(2,	0.147727051)
(3,	0.209310035)
(4,	0.285274099)
(5,	0.23817696)
        }; 
    \end{axis}
\end{tikzpicture}\vspace{1mm}%
}
\hspace{2mm}
\subfloat[\igt]{
\begin{tikzpicture}[scale=1]
    \begin{axis}[
        height=\columnwidth/2.7,
        width=\columnwidth/2.4,
        ylabel={\em Conversion Rate},
        xmin=0.5, xmax=5.5,
        ymin=0.1, ymax=0.5,
        xtick={1,2,3,4,5},
        xticklabel style = {font=\footnotesize},
        xticklabels={1,2,3,4,5},
        ytick={0.1},
        yticklabels={$10^{-1}$},
        ymode=log,
        log basis y={10},
        every axis y label/.style={at={(current axis.north west)},right=6mm,above=0mm},
        label style={font=\scriptsize},
        tick label style={font=\scriptsize},
    ]
        \addplot[line width=0.25mm,color=red,mark=triangle*] 
        plot coordinates {
(1,	0.20771496)
(2,	0.211420325)
(3,	0.18315847)
(4,	0.163387094)
(5,	0.290578622)
        };  
        \addplot[line width=0.25mm,color=forestgreen,mark=square] 
        plot coordinates {
(1,	0.207851473)
(2,	0.211556838)
(3,	0.196283006)
(4,	0.188915108)
(5,	0.251652788)
        }; 
    \end{axis}
\end{tikzpicture}\vspace{1mm}%
}
\vspace{-3mm}
\end{small}
\caption{Conversion probability of invitation behaviors conditioned on each \sitb-based measure  in \codmrecall.} \label{fig:codm-invite}
\vspace{-2mm}
\end{figure*}
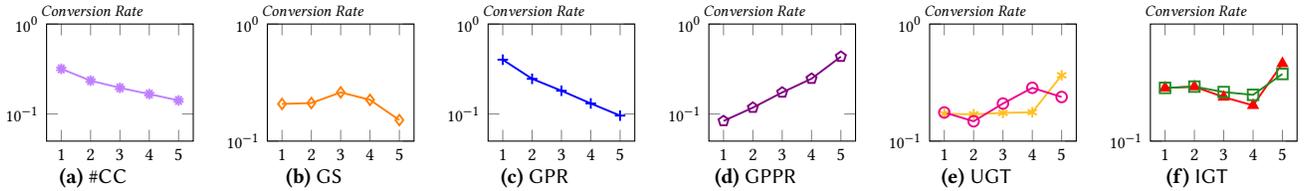

%% file: feature_importance.tex
\begin{table}[!t]
\centering
\renewcommand{\arraystretch}{1.2}
\caption{\sitb feature importance w.r.t adoption behavior.}\vspace{-2mm}
\begin{small}
\begin{tabular}{|c|c|c|c|c|c|c|}
\hline
& $\#$\cc & \gs & \cpr & \cppr & \ugt & \igt \\ \hline
\textbf{\codmrecall}  & 0.1413  & 0.1179  & 0.2291   & 0.1713    & 0.2191   & 0.1212   \\ \hline
\textbf{\lgamerecall} & 0.1561  & 0.1345  & 0.1958   & 0.1733    & 0.2139   & 0.1264   \\ \hline
\end{tabular}
\end{small}\label{tab:adopt-importance}
\vspace{-2mm}
\end{table}

\begin{table}[!t]
\centering
\renewcommand{\arraystretch}{1.2}
\caption{\sitb feature importance w.r.t invitation behavior.}\vspace{-2mm}
\begin{small}
\begin{tabular}{|c|c|c|c|c|c|c|}
\hline
& $\#$\cc & \gs & \cpr & \cppr & \ugt & \igt \\ \hline
\textbf{\codmrecall}  & 0.1404  & 0.0809  & 0.2081   & 0.2221    & 0.2271   & 0.1214   \\ \hline
\textbf{\lgamerecall} & 0.1522  & 0.1002  & 0.2540   & 0.1819    & 0.1907   & 0.1210   \\ \hline
\end{tabular}
\end{small}\label{tab:invite-importance}
\vspace{-2mm}
\end{table}

%% file: others.tex
\section{Additional Related Works}\label{sec:related}
This part briefly reviews other \tfc measures and the related sociological theory.
In particular, 
prior \tfc measures can be explained by either \textit{selection} or \textit{influence} mechanism in the homophily principle~\cite{easley2010networks}.
The selection indicates that people tend to form friendships with others with similar characteristics. In contrast, the influence can be treated as the inverse of selection, claiming that people may modify characteristics to conform to their friends. We categorize \tfc measures based on the taxonomy of selection and influence.

Regarding the selection, conventional \tfc measures utilize the variant based on common neighborhood, \eg Adamic/Adar score~\cite{adamic2003friends}. Furthermore, the selection can be measured by the similarity between node embeddings, \eg \cite{roweis2000nonlinear,tang2015line,wang2016structural,perozzi2014deepwalk,grover2016node2vec,yang13homogeneous}, which essentially rely on the factorization of a matrix of the \tfc measure as mentioned in Section~\ref{sec:exist-measure}~\cite{qiu2018network}.
Regarding social influence, the related measures mainly focus on the local structure of a user. Burt~\cite{burt1987social} proposes to model social influence from cohesion and structural equivalence perspectives. In particular, cohesion describes the direct influence measured by tie strength and the number of influenced friends~\cite{granovetter1978threshold,fang2014modeling}. The structural equivalence describes the indirect influence between users and is measured by tie embeddedness~\cite{easley2010networks}.
Recently, many learning-based methods~\cite{gomez2011uncovering,sankar2020inf, li2020chassis} are proposed to predict the influence probability based on historical cascades, among which~\cite{li2020chassis} leverages another psychological concept called conformity to measure the social influence.

In contrast, \sit is a group-level psychological framework, in which each dimension can reflect either selection or influence in the homophily principle. For instance, the factors of self-stereotyping and in-group homogeneity are coherent with the selection mechanism. Therefore, \sit can also provide theoretical support for the aforementioned measures and their integration, \eg \cite{sankar2020inf}.

\section{Conclusions}
In this work, we propose measures for friendship closeness based on the social identity theory, a fundamental concept in social psychology, and conduct extensive experiments and analysis on 3 Tencent's online gaming datasets.
Compared with 8 state-of-the-art methods, our proposal achieves the highest effectiveness on both behavior prediction and target recommendation tasks. Regarding future works, we will employ presented measures for other scenarios, \eg modeling the cascade of social influence and suggesting strangers to enrich users' friendships in the who-to-follow service.